\documentclass[sigconf]{acmart}

\newcommand{\noop}[1]{}

\AtBeginDocument{%
  \providecommand\BibTeX{{%
    \normalfont B\kern-0.5em{\scshape i\kern-0.25em b}\kern-0.8em\TeX}}}

\setcopyright{acmcopyright}
\copyrightyear{2020}
\acmYear{2020}

\acmConference[-]{-}{-}{Eindhoven, Netherlands}

\begin{document}

\title{GGDs: Graph Generating Dependencies}

 \author{Larissa C. Shimomura}
 \email{l.capobianco.shimomura@tue.nl}
 \affiliation{
   \institution{Eindhoven University of Technology}
   \city{Eindhoven}
   \country{Netherlands}
 }

 \author{George Fletcher}
 \email{g.h.l.fletcher@tue.nl}
 \affiliation{
   \institution{Eindhoven University of Technology}
   \city{Eindhoven}
   \country{Netherlands}
 }

 \author{Nikolay Yakovets}
 \email{hush@tue.nl}
 \affiliation{
   \institution{Eindhoven University of Technology}
   \city{Eindhoven}
   \country{Netherlands}
 }



\begin{abstract}
We propose 
Graph Generating Dependencies (GGDs), a new class of dependencies for property graphs.
Extending the expressivity of state of the art constraint languages, GGDs can express both tuple- and equality-generating dependencies on property graphs, both of which find broad application in graph data management.  We provide the formal definition of GGDs, analyze the validation problem for GGDs, and demonstrate the practical utility of GGDs.
\end{abstract}




\maketitle

\section{Introduction}
\label{sec:intro}

Constraints play a key role in data management research, e.g., in the study of data quality, data integration and exchange, and query optimization 
\cite{Barcelo0R13,Bohannon2007,2012Fan,Fan19,Fan2019a,Fan2016,FrancisL17,IlyasC19}. 
As graph-structured data sets proliferate in domains such as social networks, biological networks and knowledge graphs, the study of graph dependencies is also of increasing practical interest \cite{Bonifati2018,Fan19}. This raises new challenges as graphs are typically schemaless, unlike relational data. 

Recently, different classes of dependencies for graphs have been proposed such as Graph Functional Dependencies (GFDs~\cite{Fan2016}), Graph Entity Dependencies (GEDs~\cite{Fan2019a}) and Graph Differential Dependencies (GDDs~\cite{Kwashie2019}). 
However, these dependencies focus on generalizing functional dependencies (i.e., variations of {\em equality}-generating dependencies) and cannot capture {\em tuple}-generating dependencies (TGDs) for graph data \cite{Fan19}.
As an example, we might want to enforce the constraint on a human resources graph that ``if two {\em people} vertices have the same {\em name} and {\em address} property-values and they both have a {\em works-at} edge to the same {\em company} vertex, then there should be a {\em same-as} edge between the two people.''  This is an example of a TGD on graph data, as satisfaction of the constraint requires the existence of an edge (i.e., the {\em same-as} edge), and when not satisfied, we repair the graph by generating {\em same-as} edges where necessary.
TGDs are important for many applications, e.g., for entity resolution during data cleaning and integration \cite{2012Fan,IlyasC19}.  

Indeed, TGDs arise naturally in graph data management applications.  Given the lack of
TGDs for graphs in the current study of graph dependencies, we propose a new
class of graph dependencies called Graph Generating Dependencies (GGDs) which
fully supports TGDs for property graphs (i.e., TGDs for graphs where vertices and edges can have associated property values, such as names and addresses in our example above -- the most common data model in practical graph data management systems) and generalizes earlier graph
dependencies.  Informally, a GGD expresses a constraint between two (possibly)
different graph patterns enforcing relationships between property values and
topological structure.

In this short paper, we formally define GGDs, analyze the validation
problem for GGDs, and illustrate the utility of GGDs for the entity resolution problem.
We conclude the paper with indications for further study of GGDs.

\section{Related Work}
\label{sec:related}

We place GGDs in the context of relational and graph dependencies.

\textit{Relational data dependencies.} 
The classical Functional Dependencies (FDs) have been 
widely studied and extended for contemporary applications in data management.
The most related for GGDs in the state of the art are the Conditional Functional Dependencies (CFDs~\cite{Bohannon2007,2012Fan}) and the Differential Dependencies (DDs~\cite{Song2011}). CFDs were proposed for data cleaning tasks where the main idea is to enforce an FD only for a set of tuples specified by a condition, unlike the original FDs in which the dependency holds for the whole relation. 
The DDs extend the FDs by specifying looser constraints according to user-defined distance functions between attribute values.


\textit{Graph dependencies.} 
Previous work in the literature focused on defining FDs for RDF data and TGDs for graph data exchange and eliminating redundancy in RDF\cite{Barcelo0R13,Calvanese2014,FrancisL17,Pichler2010}. 
Most closely related to GGDs are the graph  functional dependencies (GFDs), graph entity dependencies (GEDs), 
and graph differential dependencies (GDDs) \cite{Fan2016,Fan2019a,Kwashie2019}.
The GFDs are formally defined as a pair $(Q[\overline{x}], X \rightarrow Y)$ in which $Q[\overline{x}]$ is a graph pattern that defines a topological constraint while $X, Y$ are two sets of literals that define the property-value functional dependencies of the GFD. 
Since graph data is usually schemaless, the property-value dependency is defined for the vertex attributes present in the graph pattern. 
The GEDs subsume the GFDs and can express FDs, GFDs, and EGDs. Besides the property-value dependencies present in the GFDs, GEDs also carry special id literals to enable identification of vertices in the graph pattern.
The GDDs extend the GEDs by introducing distance functions instead of equality functions, similar to the DDs for relational data but defined over a topological constraint expressed by a graph pattern. 
Similar to the definition of our proposed GGDs, the Graph Repairing Rules (GRRs~\cite{Cheng2018}) were proposed as an automatic repairing semantics for graphs. The semantics of a GRR is: given a source graph pattern it should be repaired to a given target graph pattern.
The graph-pattern association rules (GPARs~\cite{Fan2015}) according to \cite{Fan19} is a specific case of TGDs and has been applied to social media marketing. A GPAR is a constraint of the form $Q(x,y) \Rightarrow q(x,y)$ which states that if there exists an isomorphism from the graph pattern $Q(x,y)$ to a subgraph of the data graph, then an edge labeled $q$ between the vertices $x$ and $y$ is likely to hold.

The main differences of our proposed GGDs compared to previous works are the use of differential constraints (on both source and target side), edges are treated as first-class citizens in the graph patterns (in alignment with the property graph model), and the ability to entail the generation of new vertices and edges (see Section \ref{sec:definition} for details). 
With these new features of the GGDs, we can encode relations between two graph patterns as well as the (dis)similarity between its vertices and edges properties values. 
In general, GGD is the first constraint formalism for \emph{property graphs} supporting both EGDs and TGDs, as well as DDs for property values.

\section{Preliminaries}
\label{sec:preliminaries}
We first summarize standard notation and concepts 
\cite{Fan2019a,Song2011,Bonifati2018}.
Let $O$ be a set of objects, $L$ be a finite set of labels, $K$ be a set of property keys, and $N$ be a set of values. We assume these sets to be pairwise disjoint. 


A {\bf property graph} is a structure $(V,E,\eta, \lambda, \nu)$ where
\begin{itemize}
    \item $V \subseteq O$ is a finite set of objects, called vertices;
    \item $E \subseteq O$ is a finite set of objects, called edges;
    \item $\nu: E \rightarrow V \times V$ is function assigning to each edge an ordered pair of vertices;
    \item $\lambda: V \cup E \rightarrow P(L)$ is a function assigning to each object a finite set of labels (i.e., $P(S)$ denotes the set of finite subsets of set $S$). 
    Abusing the notation, we will use $\lambda_v$ for the function assigning labels to vertices and $\lambda_e$ for the function that assigns labels to the edges; and
    \item $\nu: (V \cup E) \times K \rightarrow N$ is partial function assigning values for properties/attributes to objects, such that the object sets $V$ and $E$ are disjoint (i.e., $V \cap E = \emptyset$) and the set of domain values where $\nu$ is defined is finite.
\end{itemize}

A {\bf graph pattern} is a directed graph $Q[\overline{x}] = (V_Q, E_Q, \lambda_Q)$ where $V_Q$ and $E_Q$ are finite sets of pattern vertices and edges, respectively, and $\lambda_Q$ is a function that assigns a label $\lambda_Q(u)$ to each vertex $u \in V_Q$ or edge $e \in E_Q$. 
Abusing notation, we use ${\lambda_v}_Q$ as a function to assign labels to vertices and ${\lambda_e}_Q$ to assign labels to edges. 
Additionally, $\overline{x}$ is a list of variables that include all the vertices in $V_Q$ and edges in $E_Q$. 

We say a label $l$ {\bf matches} a label $l'\in L$, denoted as $l \asymp l'$, if $l \in L \text{ and } l = l'$ or $l =$`-' (wildcard) . 
A match denoted as $h[\overline{x}]$ of a graph pattern $Q[\overline{x}]$ in a graph G is a homomorphism of $Q[\overline{x}]$ to G such that for each vertex $u \in V_Q, {\lambda_v}_Q(u) \asymp {\lambda_v}(h(u))$; and for each edge $e = (u,u') \in E_Q$, there exists an edge $e' = (h(u), h(u'))$ and ${\lambda_e}_Q(e) \asymp {\lambda_e}(e')$.

A {\bf differential function}  $\phi[A]$ on attribute $A$ is a constraint of difference over $A$ according to a distance metric \cite{Song2011}. Given two tuples $t_1,t_2$ in an instance I of relation R, $\phi[A]$ is true if the difference between $t_1.A$ and $t_2.A$ agrees with the constraint specified by $\phi[A]$, where $t_1.A$ and $t_2.A$ refers to the value of attribute $A$ in tuples $t_1$ and $t_2$, respectively. We use the differential function idea to define constraints in GGDs.

\section{GGD: Syntax and Semantics}
\label{sec:definition}

A {\bf Graph Generating Dependency} (GGD) is a dependency of the form \[ Q_s[\overline{x}], \phi_s \rightarrow Q_t[\overline{x},\overline{y}],\phi_t\] where:
\begin{itemize}
    \item $Q_s[\overline{x}]$ and $Q_t[\overline{x},\overline{y}]$ are graph patterns, called \textbf{source} graph pattern and \textbf{target} graph pattern, respectively;
    \item  $\phi_s$ is a set of differential constraints defined over the variables $\overline{x}$ (variables of the graph pattern $Q_s$); and
    \item $\phi_t$ is a set of differential constraints defined over the variables $\overline{x} \cup \overline{y}$, in which $\overline{x}$ are the variables of the source graph pattern $Q_s$ and $\overline{y}$ are any additional variables of the target graph pattern $Q_t$.
\end{itemize}

A differential constraint in $\phi_s$ on $[\overline{x}]$ (resp., in $\phi_t$ on $[\overline{x},\overline{y}]$) is a constraint of one of the following forms \cite{Kwashie2019,Song2011}:
\begin{enumerate}
    \item $\delta_A(x.A,c) \le t_A$ 
    \item $\delta_{A_1A_2}(x.A_1, x'.A_2) \le t_{A_1A_2}$
    \item $x = x'$ or $x \neq x'$
\end{enumerate}
where $x, x' \in \overline{x}$ (resp. $\in \overline{x} \cup \overline{y}$) for $Q_s[\overline{x}]$ (resp. for $Q_t[\overline{x},\overline{y}]$), $\delta_A$ is a user defined similarity function for the property $A$ and $x.A$ is the property value of variable $x$ on $A$, $c$ is a constant of the domain of property $A$ and $t_A$ is a pre-defined threshold. 
The differential constraints defined by (1) and (2) can use the operators $(=, <, >, \le, \ge, \neq)$. The user-defined distance function $\delta_A$ can be, for example, an edit distance when $A$ is a string or the difference between two numerical values. 

The constraint (3) $x = x'$ states that $x$ and $x'$ are the same entity (vertex/edge) and can also use the inequality operator stating that $ x \neq x'$. Since the pattern variables $\overline{x}$ in $Q_s$ (resp. $\overline{x},\overline{y}$ in $Q_t$) includes both vertices and edges, this allows to match vertex-vertex variables, edge-edge and vertex-edge variables. 

\begin{figure}
  \centering
  \includegraphics[width=.78\linewidth]{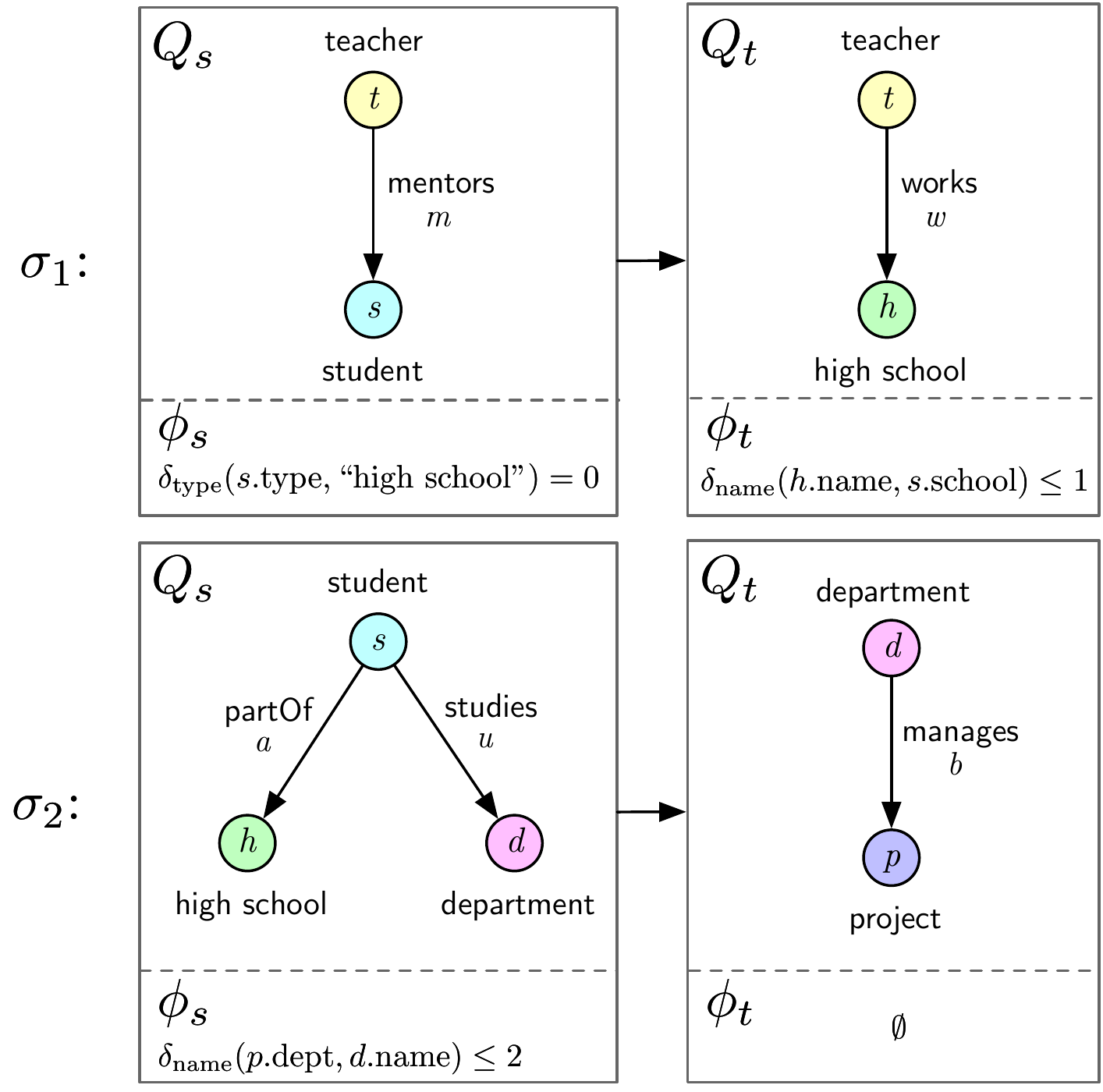}
  \caption{Example GGDs.}
  \label{fig:firstExample}
\end{figure}

\textbf{Example 1} (GGD $\sigma_1$ in \autoref{fig:firstExample}).
Here, $\sigma_1$ implies that for the matches of the source graph pattern $Q_s$, if the student type is ``high school'' then there exists a target graph pattern $Q_t$, in which the same matched vertex for teacher has an edge labelled `works' to a `high school' vertex in which the difference/(dis)similarity between the high school name and the student school name should be less than or equal to $1$.  

\textbf{Example 2} (GGD $\sigma_2$ in \autoref{fig:firstExample}). 
According to $\sigma_2$, for the matches of $Q_s$ if the project department and the department name are (dis)similar according to the threshold ``$2$" then there exists an edge labelled ``manages" linking the department and the project (graph pattern $Q_t$).

\subsection{Semantics of GGDs} 

In order to interpret a GGD $Q_s[\overline{x}], \phi_s \rightarrow Q_t[\overline{x},\overline{y}],\phi_t$, we first specify what it means for a graph pattern match to satisfy a set of differential constraints.
Consider a graph pattern $Q[\overline{z}]$, a set of differential constraints $\phi_z$ and a match of this pattern represented by $h[\overline{z}]$ in a graph $G$. The match $h[\overline{z}]$ satisfies ($\models$) a  differential constraint $k \in \phi_z$ if:
\begin{enumerate}
    \item When $k$ is $\delta_{A}(z.A,c) \le t_{A}$ then attribute $z.A$ exists at vertex/edge $z = h(z)$ and $\delta_{A}(z.A,c) \le t_{A}$ meaning that the user defined distance (for property A) $\delta_A$ between a constant $c$ and the attribute A value of vertex/edge z is less or equal than the defined threshold $t_A$.
    \item When $k$ is $\delta_{A_1A_2}(z.A_1, z'.A_2) \le t_{A_1A_2}$ then attributes $A_1,A_2$ exist at vertex/edge $z = h(z)$ and $z' = h(z')$ and $\delta_{A_1A_2}(z.A_1, \\z'.A_2)$ $\le t_{A_1A_2}$.
    \item When $k$ is $z = z'$, then $h(z)$ and $h(z')$ refer to the same vertex/edge. 
\end{enumerate}

The match $h[\overline{z}]$ satisfies $\phi_z$, denoted as $h[\overline{z}] \models \phi_z$ if the match $h[\overline{z}]$ satisfies every differential constraint in $\phi_z$. If $\phi_z = {\emptyset}$ then $h[\overline{z}] \models \phi_z$ for any match of the graph pattern $Q[\overline{z}]$ in $G$.

Given a GGD $Q_s[\overline{x}], \phi_s \rightarrow Q_t[\overline{x},\overline{y}],\phi_t$ we denote the matches of the source graph pattern $Q_s[\overline{x}]$ as $h_s[\overline{x}]$ while the matches of the target graph pattern $Q_t[\overline{x},\overline{y}]$ are denoted by $h_t[\overline{x},\overline{y}]$ which can include the variables from the source graph pattern $\overline{x}$ and additional variables $\overline{y}$ particular to the target graph pattern $Q_t[\overline{x},\overline{y}]$. 


A GGD $\sigma = Q_s[\overline{x}], \phi_s \rightarrow Q_t[\overline{x},\overline{y}],\phi_t$ holds in a graph G, denoted as $G \models \sigma$, if and only if for every match $h_s[\overline{x}]$ of the source graph pattern $Q_s[\overline{x}]$ in $G$ satisfying the set of constraints $\phi_s$, there exists a match $h_t[\overline{x},\overline{y}]$ of the graph pattern $Q_t[\overline{x},\overline{y}]$ in $G$ satisfying $\phi_t$ such that for each $x$ in $\overline{x}$ it holds that $h_s(x) = h_t(x)$. 
In case a GGD is not satisfied, we typically fix this by \emph{generating} new vertices/edges in $G$.

\begin{figure}
  \centering
  \includegraphics[width=0.9\linewidth]{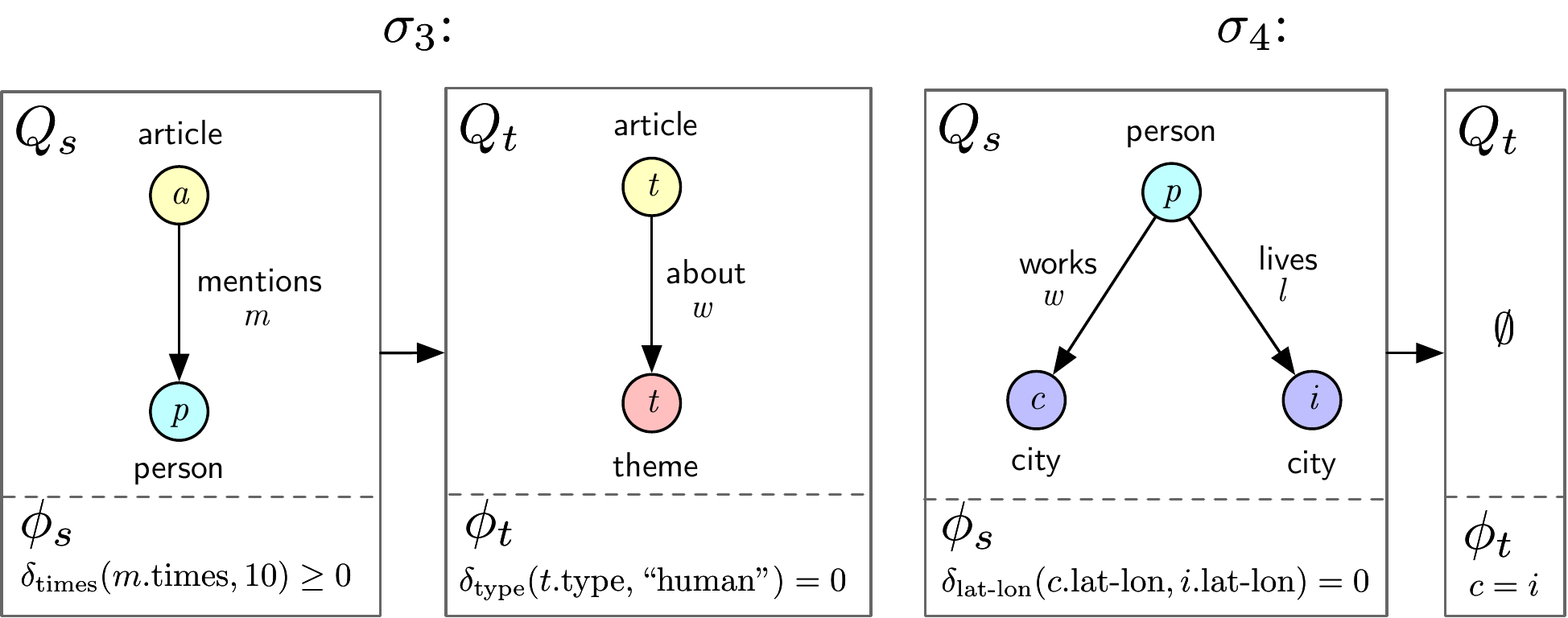}
  \caption{Example GGDs.}
  \label{fig:secondExample}
\end{figure}

\textbf{Example 3} (GGD $\sigma_3$ in \autoref{fig:secondExample}). Following the semantics of the GGDs, for every match of $Q_s$ that the number of times an article mentions a person is greater than $10$, there exists a match of $Q_t$ such that the theme type is ``human". Observe that, in this example, we use the property value of the edge variable $m$ in the differential constraint which is possible in GGDs as edges are also considered variables in the graph patterns. 

\textbf{Example 4} (GGD $\sigma_4$ in \autoref{fig:secondExample}). This GGD enforces that if the latitude and longitude coordinates of the city $c$ in which a person works and of the city $i$ in which a person lives are the same, then $c$ and $i$ should refer to the same city. Observe that in this case the target graph pattern is empty.

\begin{figure}
  \centering
  \includegraphics[width=0.8\linewidth]{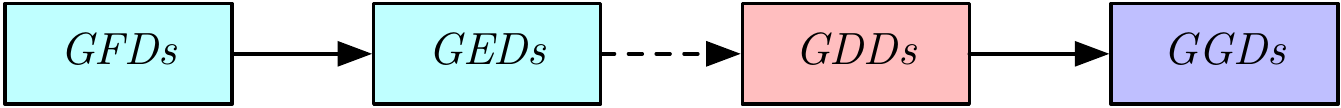}
  \caption{Expressiveness of GGDs to other graph constraints.}
  \label{fig:graphDepToGGDs}
\end{figure}

GGDs can express other graph constraints previously proposed in the literature. 
\autoref{fig:graphDepToGGDs} shows the relationship between the graph dependencies in terms of expressiveness. 
GEDs\cite{Fan19} subsumes GFDs\cite{Fan2016}, while GDDs\cite{Kwashie2019} extend GEDs by including differential constraints represented in the figure by the dashed line.
GGDs can express the GFDs, GEDs and GDDs by considering an empty target graph pattern ($Q_t[\overline{x},\overline{y}]$).
Since GEDs and GFDs only enforce equality between attributes, we can express the equality in GGDs differential constraints by using an equality operator and a threshold value $0$.

\noop{
There are three fundamental problems for GGDs:
\begin{itemize}
    \item \emph{Satisfiability} - Given a set of GGDs $\Sigma$ does there exist a non-empty graph $G$ on which all GGDs in $\Sigma$ holds, denoted as $G \models \Sigma$?
    \item \emph{Implication} - Given a set of GGDs $\Sigma$ and a GGD $\sigma$, does $\Sigma$ imply $\sigma$, (denoted by $\Sigma \models \sigma$) for every non-empty graph G that satisfies $\Sigma$? 
    \item \emph{Validation} - Given a set of GGDs $\Sigma$ and a non-empty graph $G$, does the set of GGDs $\Sigma$ holds in $G$, denoted as $G \models \Sigma$?
\end{itemize}
In this paper, we discuss the validation problem and its complexity. The Satisfiability and Implication are part of future work. 
} 

\section{Validation}
We next discuss the \textbf{validation problem} for GGDs, defined as: Given a finite set $\Sigma$ of GGDs and graph G, does $G \models \Sigma$ (i.e., $G \models \sigma$ for each $\sigma\in\Sigma$)? 
We propose an algorithm to validate a GGD $\sigma = Q_s[\overline{x}], \phi_s \rightarrow Q_t[\overline{x},\overline{y}],\phi_t$. This algorithm returns true if the $\sigma$ is validated and returns false if $\sigma$ is violated.  

We proceed as follows.  For each 
match $h_s(\overline{x})$ of the graph pattern $Q_s[\overline{x}]$ in $G$:

\begin{enumerate}
    \item Check if $h_s(\overline{x})$ satisfies the source constraints (ie., $h_s(\overline{x}) \models \phi_s$). If yes then continue.
    \item Retrieve all matches $h_t(\overline{x},\overline{y})$ of the target graph pattern $Q_t[\overline{x},\overline{y}]$ where $h_s(x) = h_t(x)$ for all $x\in\overline{x}$. If there are no such matches of the target graph pattern, return false. 
    \item Verify if $h_t(\overline{x},\overline{y}) \models \phi_t$. If there exists at least one match of the target graph pattern such that $h_t(\overline{x},\overline{y}) \models \phi_t$, then return true, else return false. 
\end{enumerate}

\noindent This process is repeated for each $\sigma \in \Sigma$. 
For each match on which $\sigma$ is violated, new vertices/edges can be generated in order to repair it (i.e, in order to make the GGD $\sigma$ valid on $G$).

We next analyse the complexity of each of the ``operations'' presented in the algorithm separately to analyse the complexity of the validation of a GGD.
Graph pattern matching queries can be expressed as conjunctive queries (CQ) \cite{Bonifati2018} which are well-known to have NP-complete evaluation complexity \cite{Pichler2011}.
The graph pattern matching problem can be solved in PTIME when the graph pattern is bounded with $k$ tree-width~\cite{Fan2016,Pichler2011}. 
To analyze the complexity of constraint checking, let $|h_s[\overline{x}]|$ be the number of matches found of the query pattern $Q_s$, $|\phi_s|$ the number of differential functions in $\phi_s$ and $f_i$ is the cost to check the differential function $i$ defined by the user, in which $0 \le i \le |\phi_s|$. The total cost for checking the differential constraints in $\phi_s$ is: $|h_t[\overline{x}]|(|\phi_s| \sum_{i=0}^{|\phi_s|} f_i)$.

For each of the matches that satisfies the differential functions in $\phi_s$, we verify the target side of the differential constraint, $Q_t[\overline{x},\overline{y}],\phi_t$. 
Assuming that the cost for checking the differential functions is tractable, 
we can show that the complexity of the validation problem of GGDs follows from the evaluation problem for classical relational tuple-generating dependencies, i.e., has $\Pi_{2}$P-complete complexity \cite{Pichler2011}. 
Pichler and Skritek have established polynomial time validation complexity for a large subclass of tgds \cite{Pichler2011}, which corresponds to graph patterns 
covering over 99\% of graph patterns
observed in practice \cite{BonifatiMT20}. 


\section{GGDs for Entity Resolution}
\label{sec:usecases}

The main novelty of the GGDs is in the generation of new vertices or edges in case a GGD is violated. Given this feature, GGDs can be applied in different scenarios. In this section, we show how GGDs can be used in solutions for entity resolution (ER). 

ER is the task of identifying and linking entities across (possibly) different data sources that refer to the same real-world entity\cite{2012Fan,IlyasC19}. 
The generation of new vertices and/or edges in case a GGD is violated gives the possibility to rewrite ER matching rules or conditions as GGDs. Towards entity resolution we can define the source graph patterns as several disjoint patterns from (possibly) different graph sources and use the target graph pattern specifications as the representation of the deduplicated graphs. 
Thus, using this approach, we can also encode more information than just vertex-to-vertex, or row-to-row in relational databases, as we consider all the information in a defined graph pattern. 

\begin{figure}
  \centering
  \includegraphics[width=0.95\linewidth]{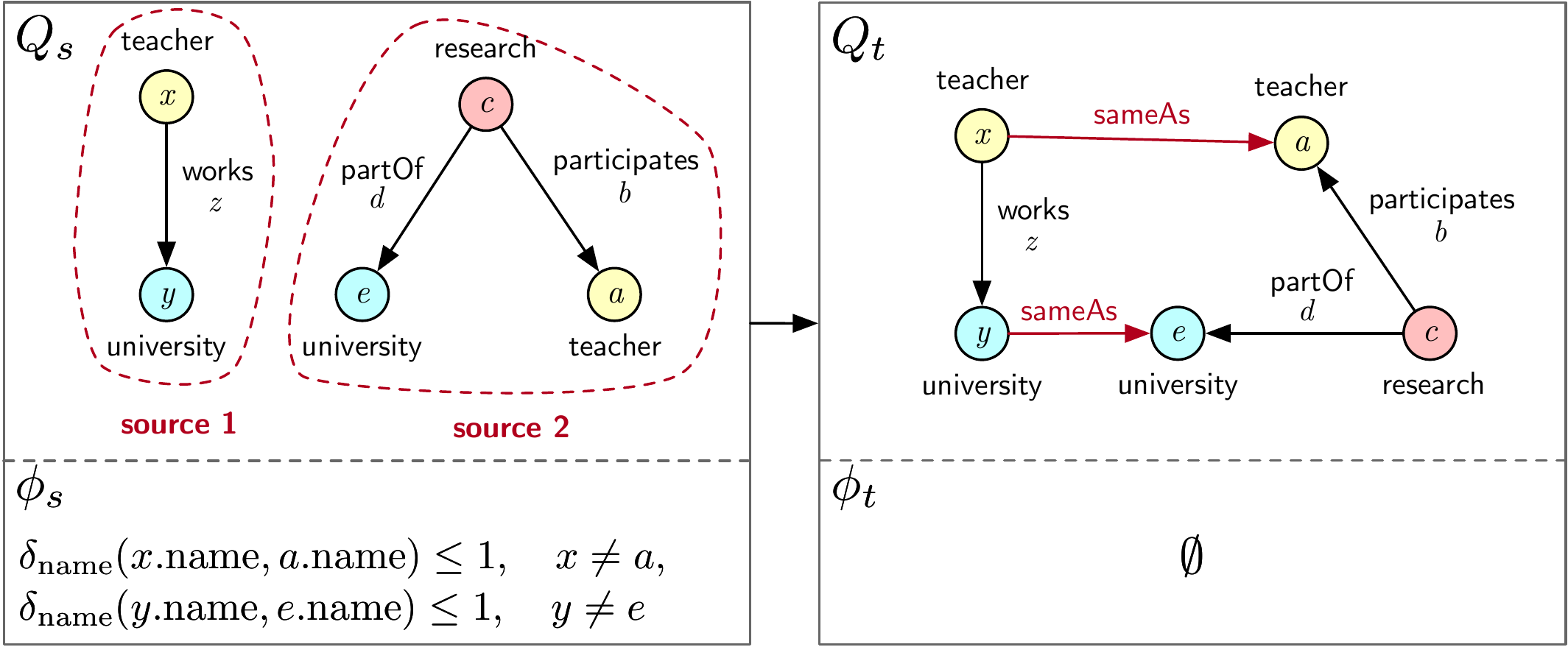}
  \caption{GGD for Entity Resolution.}
  \label{fig:exampleER}
\end{figure}

\textbf{Example 5} (\autoref{fig:exampleER}). As discussed before, the source graph pattern encodes the rules to perform entity resolution over (possibly) different graph sources. 
To perform ER, we can add links of type `sameAs' between the matched entities in the target graph pattern. These links will be generated to validate the defined GGD.  

A second interesting case in which the GGDs can be used to solve entity resolution is when two graph patterns that refer to the same real-world entity have different structures in (possibly) different sources. In this case, we can generate with the GGDs a vertex or a graph pattern that can summarize all the information of these two graph patterns (see \textbf{Example 6}). 
An advantage in using GGDs for the ER is the use of edges as variables, allowing to use the information of edge properties also in the matching rules, as it can be observed in the next example.

\begin{figure}
  \centering
  \includegraphics[width=0.85\linewidth]{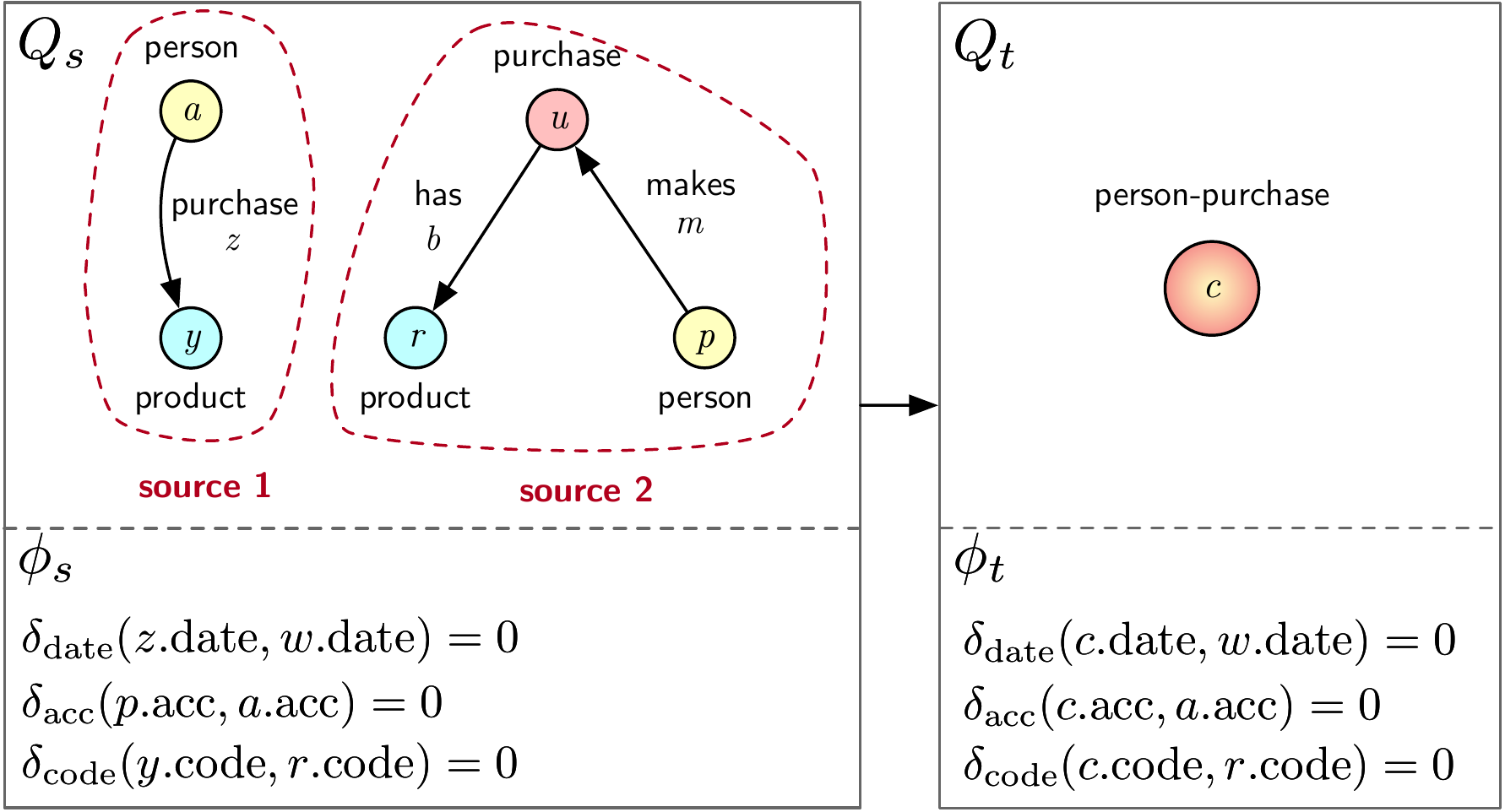}
  \caption{GGD for Entity Resolution.}
  \label{fig:exampleER2}
\end{figure}

\textbf{Example 6} (\autoref{fig:exampleER2}). In this case, we have two graph sources that model differently the same act of purchasing a product. In order to deduplicate this data, it is useful to create a vertex in the integrated graph that is able to aggregate the information that matches in both sources. 

\section{Conclusion and Future Work} 

Motivated by practical applications in graph data management,
we proposed a new class of graph dependencies called Graph Generating Dependencies (GGDs). The GGDs are inspired by the tuple- and equality-generating dependencies from relational data, where constraint satisfaction can generate new vertices and edges. 
A GGD defines a graph dependency between two (possibly) different graph patterns and the constraints over the property values are differential constraints. 
We also presented the complexity of the validation problem as well as how GGDs can be applied in the problem of ER.
 
As future work, we plan to study the satisfiability and implication problems for the GGDs, inference rules, tractable cases, the discovery of GGDs, repair of GGDs, and also further apply the GGDs to other tasks in graph data management.

\begin{acks}
 \smallskip
     \noindent {\bf Acknowledgments.} This project has received funding from the European Union's Horizon 2020 research and innovation programme under grant agreement No 825041.
\end{acks}

\bibliographystyle{ACM-Reference-Format}
\bibliography{paper}





\end{document}